\documentclass[aps,prl,amsmath,amssymb]{revtex4}
\usepackage{graphicx}
\usepackage{amsmath}
\usepackage{color}
\usepackage[normalem]{ulem}

\begin{document}

\title{Angstrom-resolution single-molecule fluorescence resonance energy transfer reveals mechanisms of DNA helicases}
\author{Wenxia Lin$^{1}$, Jianbing Ma$^{1}$, Daguan Nong$^{1}$, Chunhua Xu$^{1}$, Bo Zhang$^{2}$,
Jinghua Li$^{1}$, Qi Jia$^{1}$, Shuoxing Dou$^{1,4}$, Xuguang
Xi$^{2,3}$}
\author{Ying Lu$^{1}$} \email{yinglu@iphy.ac.cn}
\author{Ming Li$^{1,4}$} \email{mingli@iphy.ac.cn}

\affiliation{$^{1}$Beijing National Laboratory for Condensed Matter
Physics and CAS Key Laboratory of Soft Matter Physics, Institute of
Physics, Chinese Academy of Sciences, Beijing 100190, China.\\
$^{2}$College of Life Sciences, Northwest A $\&$ F University, Yangling, Shaanxi 712100, China. \\
$^{3}$LBPA, ENS de Cachan, CNRS, Universit¨¦ Paris-Saclay, F-94235 Cachan, France. \\
$^{4}$School of Physical Sciences, University of Chinese Academy of
Sciences, Beijing 100049, China.\\}

\date{\today}

\begin{abstract}
Single-molecule FRET is widely used to study helicases by detecting
distance changes between a fluorescent donor and an acceptor
anchored to overhangs of a forked DNA duplex. However, it has lacked
single-base pair (1-bp) resolution required for revealing stepping
dynamics in unwinding because FRET signals are usually blurred by
thermal fluctuations of the overhangs. We designed a nanotensioner
in which a short DNA is bent to exert a force on the overhangs, just
as in optical/magnetic tweezers. The strategy improved the
resolution of FRET to 0.5 bp, high enough to uncover the differences
in DNA unwinding by yeast Pif1 and \textit{E. coli} RecQ whose
unwinding behaviors cannot be differentiated by currently practiced
methods. We found that Pif1 exhibits 1-bp-stepping kinetics, while
RecQ breaks 1 bp at a time but sequesters the nascent nucleotides
and releases them randomly. The high-resolution data allowed us to
propose a three-parameter model to quantitatively interpret the
apparently different unwinding behaviors of the two helicases which
belong to two superfamilies.
\end{abstract}


\maketitle

Helicases are motor proteins involved in almost every aspect of
nucleic acid metabolism \cite
{Patel(1),Patel(2),Singleton(3),Lohman(4)}. When a helicase is
loaded onto one of the overhangs (i.e., the tracking strand) of a
forked DNA duplex and unwinds it, two nucleotides would be released
per base pair unwound, leading to an increase in end-to-end distance
of the overhangs. It is of great interest to know how a helicase
uses the discrete energy derived from NTP hydrolysis to unwind DNA.
The question remained unanswered for most helicases due to the lack
of proper methods that can interrogate the stepping kinetics of
helicases \cite {Lohman(5),Sikora(6),Ramanagoudr-Bhojappa(7)}.
Optical tweezers (OT) are so far the most reliable technique to
study single helicases with 0.5-bp resolution \cite
{Qi(8),Dumont(9),Cheng(10)}. However, high-resolution measurements
with OT require complicated instrumentation that is accessible to
few laboratories only. In addition, OT measures one molecule at a
time so that the throughput is usually very low. smFRET is a
high-throughput technique for helicase assays. Using wide-field
fluorescence microscopy, one can routinely record signals in
parallel from hundreds of single molecules tethered to a surface
\cite {Ha(11),Syed(12),Myong(13),Lee(14)}. To date, the resolution
of smFRET is limited to 2-3 bp when forked DNA is used as the
substrate because of the softness of the single-stranded overhangs
(Figure 1a). If a tension is exerted on the overhangs to suppress
the fluctuations, the distance between the two fluorophores can be
both stabilized and increased and the resolution of smFRET would
thus be improved. The idea has led to the combination of optical or
magnetic tweezers with smFRET in a few singe-molecule studies \cite
{Lee(15),Long(16),Kemmerich(17),Comstock(18),Graves(19)}. These
experiments, however, are even more difficult to perform because the
fluorophores are prone to bleach during operations with the
tweezers.

  In the present work, by mimicking the tweezer-enhanced smFRET, we
designed a nanotensioner with which a short DNA duplex is bent to
stretch the two overhangs of the forked DNA (Figure 1b). Due to
mismatch of their lengths, the short DNA duplex is bent to an arc,
exerting a force on the fork. The force depends on the DNA
construction. It is about 6 pN in one of our designs and the FRET
change induced by each base pair unwound is about 0.13 (Figure 1),
which is large enough to be readily recorded by many commercial
single-molecule fluorescence microscopes. In contrast, the value is
only about 0.04 when a forked DNA with free overhangs is used
(Figure 1a). We applied the nanotensioner method to assess two
helicases, namely, the yeast Pif1 and the \textit{E. coli} RecQ.
Pif1 is a prototypical member of the 5$'$ -3$'$ directed helicase
superfamily 1B that is conserved from yeast to human \cite
{Bessler(20),Bochman(21),Li(22)}. It plays critical roles in the
maintenance of telomeric DNA via catalytic inhibition of telomerase
\cite {Schulz(23),Zhou(24),Makovets(25),Paeschke(26)}. RecQ is a
member of superfamily 2 helicases \cite {Janscak(27),Bernstein(28)}.
It translocates in the 3$'$ to 5$'$ direction and contains the
conserved DEAH box motif \cite {Fairman-Williams(29)}. RecQ plays an
important role in DNA damage response, chromosomal stability
maintenance and has a vital role in maintaining genome homeostasis
\cite {Xu(30),Chu(31),Bjergbaek(32),Cobb(33)}. We did not observe
significant differences between the unwinding behaviors of the two
helicases with the forked DNA substrates as in conventional FRET
assays. With the nanotensioner, however, the difference became
obvious, enabling us to interrogate the molecular mechanisms of the
two helicases with unprecedentedly deep insight.

  The design of the nanotensioner (Figure 1b) is based on the fact
that dsDNA is semi-flexible \cite {Bustamante(34)} with a bending
modulus of $B\approx200$ pN${\times}$nm$^2$. When a short dsDNA with
a contour length $S$ is bent to an arc of radius $R$, the bending
energy is $W = BS/2R^2$. An ssDNA string of length $x =
2R\sin(S/2R)$ maintains the radius of the dsDNA arc by exerting a
force $F$ on the two ends of the dsDNA segment. The force can be
calculated according to $F=-{\delta}W/{\delta}x$, which reads as
\cite {Choi(35)}
\[
 F=\frac{BS}{R^3}\left[2\sin\left(\frac{S}{2R}\right)-\frac{S}{R}\cos\left(\frac{S}{2R}\right)\right]^{-1}.
 \tag{1}
\]
It is also the tension exerted on the ssDNA overhangs of the forked
DNA to be unwound. We attached a pair of dyes (Cy3 and Cy5) near the
junction between the ssDNA overhangs and the duplex of the forked
dsDNA. It is worth noting that, to serve our purpose of studying the
unwinding stepping kinetics of a helicase, the DNA construct should
meet the following conditions: (i) The tension on the string should
be strong enough so that the change in FRET induced by one base pair
unwound is not smaller than 0.1, which is the value required to
recognize a step by our instrument; (ii) The tension does not
decrease too much upon duplex unwinding so that the steps are
basically uniform during the experiments.

  Our calculations indicated that the DNA construct in Figure 1b will
meet the above conditions and work well when the initial length of
the string is in the range from 15 to 40 nt and the length of the
dsDNA arc ranges from 50 to 100 bp. Shown in Figure 1c is the
calculated FRET efficiency versus number of base pairs unwound for
three typical nanotensioners. The average FRET change is about 0.13
as 1 bp is unwound, which is large enough to be resolved by our
instrument. The tension decreases slightly when the string length
increases due to the newly added nucleotides (Figure 1d). In the
present work, we used a DNA nanotensioner with a 60-bp dsDNA arc and
a 30-nt ssDNA string. The DNA duplex to be unwound had a length of
40 bp, of which only ~15 bp from the junction would be effective in
the smFRET assay because of the limited measurement range of FRET
($<$10 nm).

   In order to show advantages of FRET with nanotensioners, we first
characterized the unwinding kinetics with a forked DNA substrate as
in conventional FRET assays (Figure 1a). The experiments were
performed with objective-based total-internal-reflection
fluorescence microscopy in imaging buffer composed of 25 mM Tris-HCl
(pH 7.5), 50 mM NaCl, 5 mM MgCl$_2$, 2 mM DTT and an oxygen
scavenging system ( $0.8\%$ D-glucose; 1 mg/ml glucose oxidase; 0.4
mg/ml catalase; 1 mM Trolox). Similar unwinding bursts were observed
for the both helicases in buffers containning 5 nM helicase and 20
$\mu$M ATP. Because the DNA unwinding rate can be regulated by ATP
concentration, we reduced the ATP concentration to 0.5-2 $\mu$M in
order to see possible unwinding steps \cite{Syed(12),Myong(13)}.
Unfortunately, it was not easy to identify stepping events for
either Pif1 or RecQ. The results indicated that thermal fluctuations
of the displaced ssDNA indeed blur the FRET signals. Moreover, the
fluctuations make it hard to convert the FRET changes to step sizes
unambiguously.

   We repeated the experiments to show the feasibility of the
nanotensioner method. We observed distinct unwinding steps when the
ATP concentration was reduced to 1 $\mu$M (Figure 2a). The
corresponding FRET values are in agreement with the theoretical
calculations assuming that Pif1 unwinds 1 bp at a time. We built a
histogram of the steps (Figure 2b) in the FRET range from 0.2 to 0.8
in which an approximately linear relationship exists between the
distance of the two dyes and the observed FRET value \cite
{Harada(36),Blosser(37)}. The distribution is narrow and has a peak
at ${\Delta}FRET = 0.14 \pm 0.02$. The measurements at 0.5 $\mu$M
ATP yielded similar results. We also built histograms of dwell times
before each stepping event (Figure 2c). Nonlinear least-squares
analyses indicate that these distributions are best described by a
single-exponential function, suggesting that these dwells are
governed by a single kinetic event. It is an expected behavior for
binding of a single ATP molecule before each step under limiting ATP
concentrations. The decay time ($0.32 \pm 0.08$ s) at 1 $\mu$M ATP
is about half of that ($0.58 \pm 0.06$ s) at 0.5 $\mu$M ATP,
suggesting that the dwell is dominated by the time the protein takes
to bind a single ATP. All together, these results are consistent
with a model in which the enzyme hydrolyses a single ATP at a time
and unzips the duplex with a uniform step size of 1 bp.

 In contrast to the results for Pif1 in Figure 2, the unwinding steps
of RecQ are not uniform even with the nanotensioner substrate. Shown
in Figure 3 are representative unwinding bursts of RecQ recorded at
low ATP concentrations. We observed events in which the DNA
overhangs increased in length with various increments. The
distributions of the increments, i.e., the step sizes, are composed
of multiple peaks. Beside the expected peak corresponding to
unwinding of 1 bp, the distribution at 2 $\mu$M ATP shows other
peaks corresponding to unwinding of 0.5, 1.5, 2.0, 2.5 and 3.0 bp,
respectively. Similarly, the distribution of step sizes at 5 $\mu$M
ATP also shows a few major peaks, centered at 1.0, 1.5, 2.0, 2.5,
3.0 and 3.5 bp, respectively.

  It is obvious that the unwinding kinetics of RecQ is very different
from that of Pif1. The difference is detectable only when the
resolution is better than 0.5 bp; even with a 1-bp resolution, one
may still not be able to measure the 0.5 bp steps and get only a
smeared wide peak in the histogram, hence mistakenly drawing the
conclusion that RecQ unwinds DNA with step sizes of 2-4 bp.
Irrespective of the big differences, we show in the following that
one can have a general understanding of the stepping kinetics of
Pif1, RecQ and, possibly, other helicases. To this end, we argue
that (i) a helicase breaks a base pair upon ATP hydrolysis,
generating single-stranded nucleotides, 1 nt for each tail of the
forked DNA; (ii) unwinding steps, however, are observable only when
the generated nucleotides are released; and (iii) in general, it is
not necessary that a nucleotide be released concomitantly with its
generation because it may stick to some domain of the helicase
through, e.g., electrostatic interaction and/or hydrogen bonds
\cite{Cheng(10),Newman(38)}. In other words, the helicase may
sequester the nascent nucleotides and then release them after a
random number of 1-bp opening events (Figure 4). The very general
depiction involves three independent parameters, i.e., a base-pair
breaking rate  $k_b$, a 3$'$-tail releasing rate $k_{r1}$ and a
5'-tail releasing rate $k_{r2}$. The base-pair breaking rate $k_b$
of a helicase is regulated by the concentration of ATP and can be
derived from the unwinding rate versus ATP curve. There is, however,
no direct way to calculate the two tail-releasing rates $k_{r1}$ and
$k_{r2}$. They can be estimated by using Monte Carlo simulations
\cite{Monto Carlo} to reconstruct the histograms of dwell times and
step sizes according to the kinetic model sketched in Figure 4.

  In our Monte Carlo simulations we made an assumption that the two
tail-releasing rates $k_{r1}$ and $k_{r2}$ increase with the number
of nucleotides sequestered by the helicase. This is necessary
because, otherwise, when $k_b$ became much larger than $k_{r1}$ and
$k_{r2}$ at high ATP concentrations, the number of nucleotides held
by the helicase might become too long to be true. A few physical
factors may underlie the assumption. Plausible ones include the
reduction of entropy due to confinement of the nucleotides and/or
the increase in energy due to accumulation of the negative charges
of DNA on the surface of the protein. For simplicity, we assume that
the extra energy is proportional to the number of nucleotides held
by the helicase. The tail-releasing rates can hence be written as
$k_{ri}=k_{ri}^0\exp[\alpha(n-1)]$, where $i = 1$ or 2 and
$k_{ri}^0$ is the rate when $n = 1$. Using the values of the
breaking rate $k_b$ from literature \cite{Klaue(39)}, 3.3 s$^{-1}$
at 2 $\mu$M ATP and 6.8 s$^{-1}$ at 5 $\mu$M ATP, the reconstructed
histograms for RecQ resemble the measured ones when the following
parameters are used: $k_{r1}^{0}\approx k_{r2}^{0} = 0.3$ s$^{-1}$
and $\alpha$ = 0.7 (Figures. 4b and c). We can also use the Monte
Carlo simulation to reconstruct the histograms for Pif1. The
nucleotide releasing rates of Pif1 must be, however, at least one
order of magnitude higher than the base-pair breaking rate in order
to have a good fit. This is equivalent to say that the nucleotides
are released immediately after they are generated. As a consequence,
the observable unwinding step sizes is 1 bp and the dwell time
distributions are exponential with characteristic dwell times
depending on the concentration of ATP (Figure 2c). Taken together,
the kinetic model sketched in Figure 4 applies to the both helicases
with apparently different unwinding behaviors. In addition, a few
more simulations with different parameters (data not shown) implied
that a histogram of dwell times does not necessarily follow a simple
function; it may even not be monotonic under certain conditions.

  smFRET has become the technique of choice to study helicases.
However, to the best of our knowledge, it has not yet been able to
resolve 1-nt step size until now. On one hand, the unprecedented
resolution provided by the presented nanotensioner approach enabled
us to reveal the details of helicase-catalyzed DNA unwinding that
are not easy to study with conventional smFRET method. On the other
hand, the high-resolution data also allowed us propose a unified
molecular mechanism for the two helicases that belong to two
different superfamilies with apparently different unwinding
behaviors, implying that many helicases might be more foundamentally
correlated.

{\it Acknowledgements.} This work was supported by a National
Science Foundation of China (Grant No. 11674382 (to Y.L.), 11574382
(to M.L.) and 11574381 (to CHX)) and by the Key Research Program of
Frontier Sciences, CAS (Grant No. QYZDJ-SSW-SYS014 (to M.L.)).

\begin{figure}[h]
\includegraphics[width=8cm] {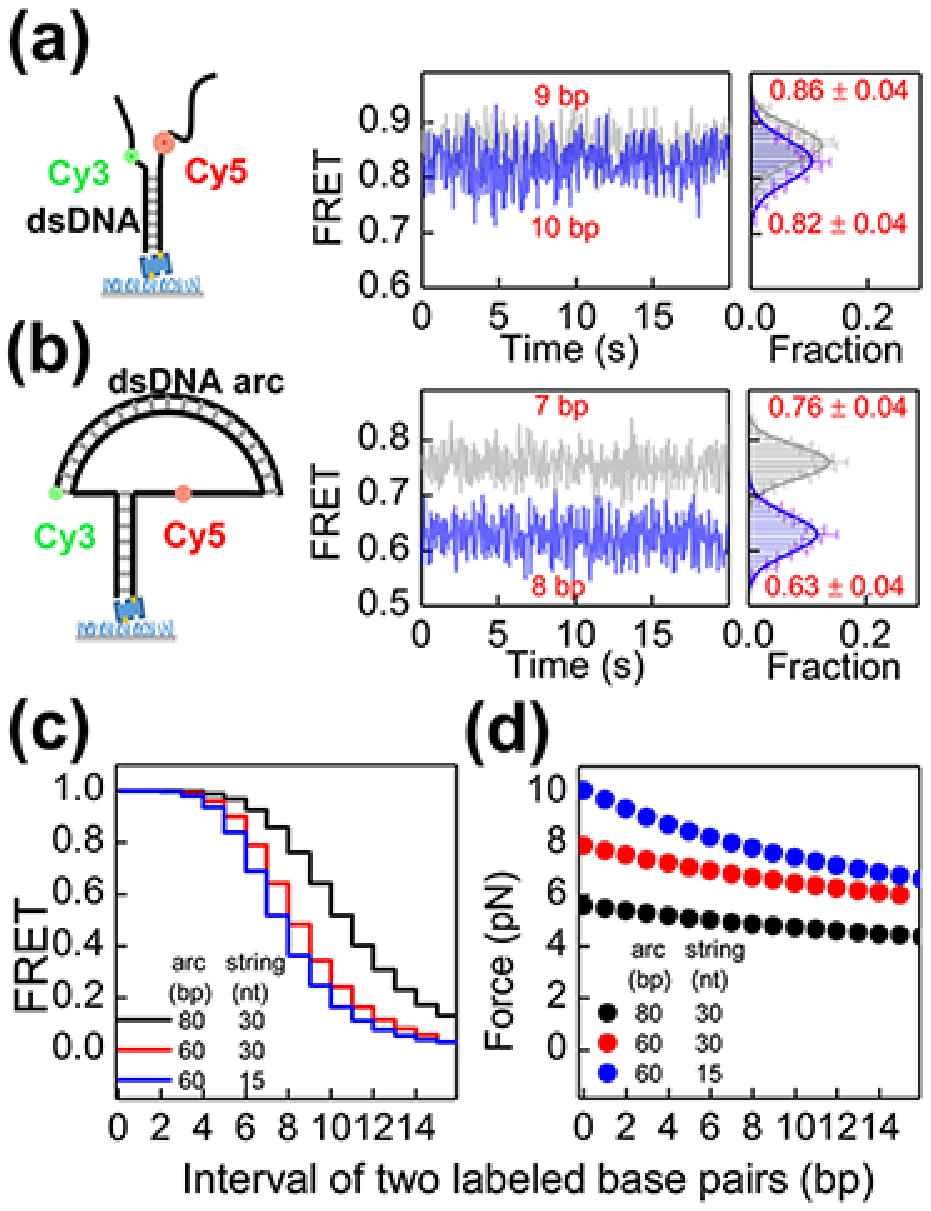}
\caption{(Color online). Design of DNA nanotensioners. (a) Free
overhangs of a forked DNA fluctuate to blur the FRET signals. The
mean FRET efficiency decreases by ~0.04 when the Cy3-Cy5 pair is
moved from the 9$^{th}$ to the 10$^{th}$ nucleotide away from the
fork. (b) In a nanotensioner, the two ends of a short dsDNA are
connected to two ssDNA overhangs of a forked DNA. The dsDNA is bent
to exert a force on the fork. The mean FRET efficiency decreases by
~0.13 when the Cy3-Cy5 pair is moved from the 7$^{th}$ to the
8$^{th}$ nucleotide. (c, d) Calculated FRET efficiency (c) and
tension (d) as a function of base pairs unwound for various
nanotensioners.} \label{schematic}
\end{figure}
\end{titlepage}
\begin{titlepage}
\begin{figure}[h]
\includegraphics[width=8cm] {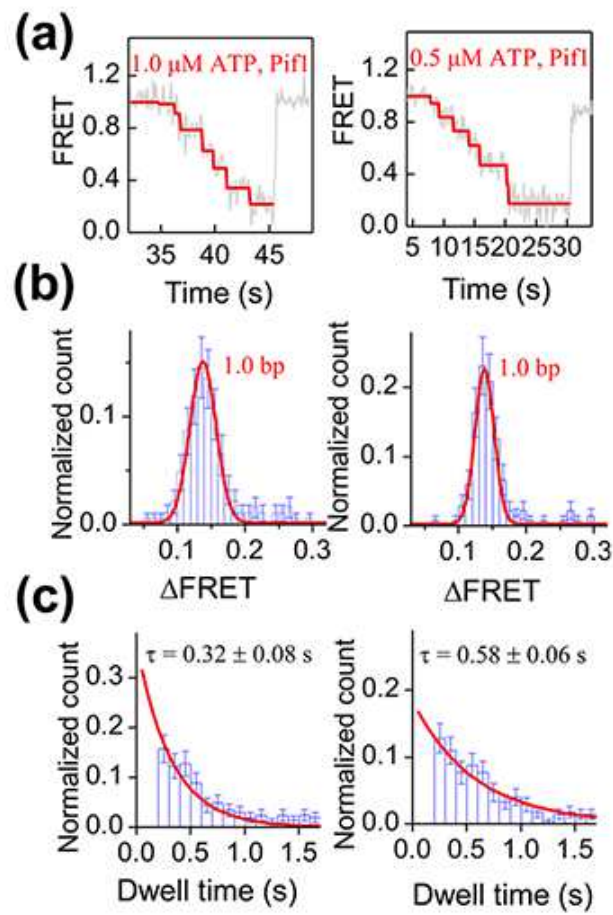}
\caption{(Color online). Pif1 unwinds 1 bp at a time in the assay
with nanotensioner. (a) Typical FRET traces. (b) Histogram of FRET
jumps ($\Delta$FRET). The red lines represent Gaussian fittings. (c)
Histogram of dwell times before each stepping event. The red lines
represent single-exponential fittings. The statistics are from $>$
150 traces at each ATP. The error bars denote SD. The errors in the
fits are SEM. } \label{schematic}
\end{figure}
\end{titlepage}
\begin{titlepage}
\begin{figure}[h]
\includegraphics[width=8cm] {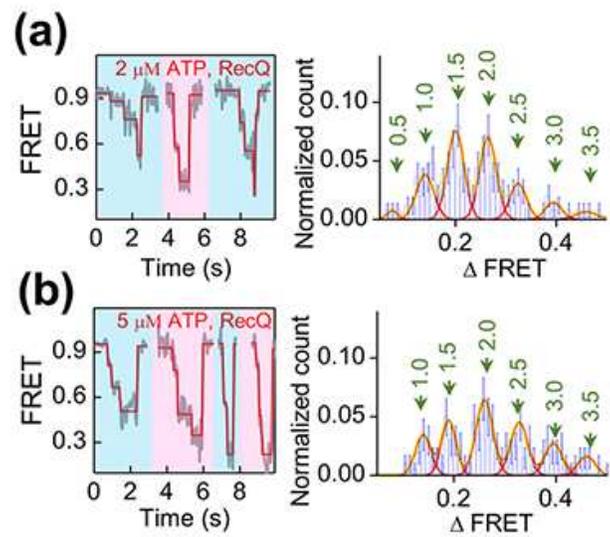}
\caption{\label{fig:wide}(Color online). Non-uniform stepwise
unwinding by RecQ in the assay with nanotensioner. (a) Typical FRET
traces and the corresponding distribution of FRET jumps
($\Delta$FRET) at 2 $\mu$M ATP. (b) Similar data recorded at 5
$\mu$M ATP. The individual traces are shifted horizontally for
clarity. The red lines in the histogram represent Gaussian fittings.
The step sizes converted from the FRET jumps are indicated on tops
of the peaks. The data are from $>$ 200 traces at each ATP. The
error bars denote SD.} \label{schematic}
\end{figure}
\end{titlepage}
\begin{titlepage}
\begin{figure}[h]
\includegraphics[width=8cm] {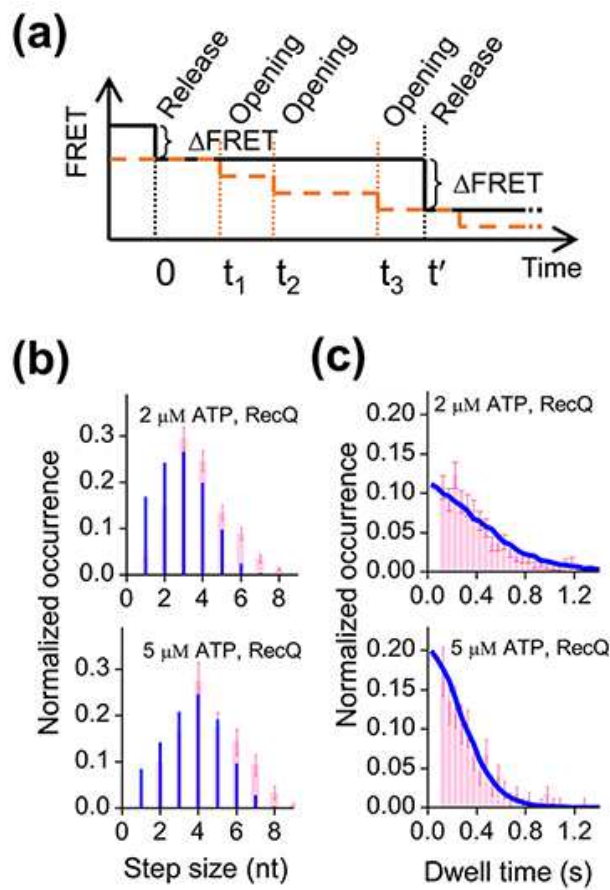}
\caption{(Color online). Stepping kinetics of helicase-catalyzed DNA
unwinding. (a) Schematic time course of observable FRET signal
(black line). At $t = 0$ and $t'$, FRET jumps appear because of the
release of the sequestered nascent nucleotides on the 3'- or
5'-strand. The orange dashed line represents the expected time
course of the FRET signal if nucleotide release is synchronous with
base pair opening, as for Pif1. In the time duration from $t = 0$ to
$t'$, base pair opening occurs three times at $t_1$, $t_2$ and
$t_3$, respectively. (b, c) Comparison between the Monte Carlo
simulated (blue) and the experimentally measured (pink)
probabilities of step sizes and dwell times of RecQ. The
experimental probabilities in b are peak values in the histograms in
Figure 3.} \label{schematic}
\end{figure}
\end{titlepage}

\end{document}